\documentclass[conference]{IEEEtran}
\IEEEoverridecommandlockouts
\usepackage{cite}
\usepackage{amsmath,amssymb,amsfonts}
\usepackage{algorithmic}
\usepackage{graphicx}
\usepackage{textcomp}
\usepackage{xcolor}
\usepackage{orcidlink}
\usepackage{booktabs}
\usepackage{mdframed}

\def\BibTeX{{\rm B\kern-.05em{\sc i\kern-.025em b}\kern-.08em
    T\kern-.1667em\lower.7ex\hbox{E}\kern-.125emX}}

\newcommand{\etal}[0]{~\textit{et al.}}

\makeatletter
\newcommand{\linebreakand}{%
  \end{@IEEEauthorhalign}
  \hfill\mbox{}\par
  \mbox{}\hfill\begin{@IEEEauthorhalign}
}
\makeatother

\begin{document}

\title{Large Language Models Have Unreliable Understanding of Software Engineering Terminology
}

\author{\IEEEauthorblockN{1\textsuperscript{st} Huzaifa Ejaz}
\IEEEauthorblockA{\textit{Faculty of Computer Science and Mathematics} \\
\textit{University of Passau}\\
Passau, Germany \\
ejaz01@ads.uni-passau.de}
\and
\IEEEauthorblockN{2\textsuperscript{nd} Fabian C. Peña \orcidlink{0009-0008-2249-7990}}
\IEEEauthorblockA{\textit{Faculty of Computer Science and Mathematics} \\
\textit{University of Passau}\\
Passau, Germany \\
fabiancamilo.penalozano@uni-passau.de}
\linebreakand
\IEEEauthorblockN{3\textsuperscript{rd} Steffen Herbold \orcidlink{0000-0001-9765-2803}}
\IEEEauthorblockA{\textit{Faculty of Computer Science and Mathematics} \\
\textit{University of Passau}\\
Passau, Germany \\
steffen.herbold@uni-passau.de}
}

\maketitle

\begin{abstract}
Large Language Models (LLMs) are increasingly used in software engineering (SE), yet there is no systematic study that determines to which degree these LLMs actually understand standardized SE terminology. Lack of such understanding can lead to miscommunication and misunderstanding, both by LLMs consuming text but also by human-developers acting on LLM-generated text. Within this paper, we investigate to which degree state-of-the-art LLMs are able to identify whether definitions from the \textit{ISO/IEC/IEEE 24765:2017 Systems and Software Engineering — Vocabulary} are correct. We prompt LLMs both with correct definitions, as well as systematically falsified definitions. The falsifications are both semantic (substitution of key terms) and structural (removing critical information). We measure both classification accuracy and whether reasoning tokens generated by the LLMs make sense with respect to understanding the definition. While most LLMs detect falsified definitions with high accuracy, they also reject many correct definitions, indicating a systematic rejection bias rather than genuine discriminative understanding. Explicit reasoning does not consistently improve results and may even hinder performance through over-thinking. Our work demonstrates that while the performance of LLMs (including their agentic use) in many SE tasks is impressive, there are still fundamental issues to understand how this will impact SE, including the consistent use of terminology. 
\end{abstract}


\section{Introduction}

Large Language Models (LLMs) have become deeply integrated into the software engineering (SE) practice. For example, a systematic literature review by Hou\etal~\cite{hou2024llm4se}, analyzing 395 studies, found LLMs applied across 85 distinct SE tasks spanning the entire software development lifecycle, from requirements engineering and code generation to testing, maintenance, and documentation. The rapid adoption of these LLMs reflects their demonstrated ability to process and generate both natural language and code with increasing fluency~\cite{10.1145/3605943, 10.1145/3747588}. The strong and ever-increasing performance of frontier LLMs in benchmarks such as SWE-bench~\cite{swebench} or Terminal-bench~\cite{terminalbench} notwithstanding, we still lack a principled understanding of the SE knowledge that leads to this success. 

Within this paper, we study one specific, low-level SE capability: understanding the correctness of definitions from the SE context. This Natural Language Understanding (NLU) task is critical, since SE is a domain that relies heavily on precisely defined concepts to facilitate communication between different actors~\cite{calero2006ontologies}. Concepts such as verification and validation, or fault, failure, and error, are not interchangeable: each refers to a distinct phenomenon with specific implications for how software is developed, tested, and assessed. A LLM that conflates these terms, or accepts a subtly incorrect definition as valid, is also likely to misuse such terms later. Such misuse could silently lead to a multitude of issues, e.g., misunderstanding inputs and, therefore, generating faulty outputs, but also generating software artifacts that misuse terminology, possibly leading to communication issues with downstream actors (both LLM and human), that consume these artifacts. 

\begin{description}
    \item[\textbf{RQ}] Do LLMs understand SE terminology on an expert level and such that they judge the correctness of definitions?
\end{description}

The starting point of our investigation into this research question is two assumptions about the training of frontier LLMs: (1) they are trained with access to high-quality data, including standardization documents; and (2) they are trained with diverse data from the internet (e.g., developer forums, informal documentation) where definitions are not always used rigorously and sometimes also in plausible-sounding, but incorrect ways. Based on these reasonable assumptions, we theorize that LLMs are very good at identifying correct definitions as such, since they were provided during training and the known memorization capabilities of LLMs~\cite{carlini2022quantifying}, also demonstrated within the SE domain~\cite{casparistudying}, should facilitate this. However, we believe that this is somewhat offset by the data diversity, which leads LLMs to accept too many uses of terms as correct, leading to problems with the identification of falsified definitions.  Thus, we derive the following two hypotheses: 

\begin{description}
    \item[\textbf{H1}] LLMs are near perfect when asked to judge whether a correct definition of a SE term is correct.
    \item[\textbf{H2}] LLMs struggle when asked to judge whether a SE definition with systematically induced subtle falsification is correct. 
\end{description}

We study these hypotheses using a confirmatory experiment. We use the \textit{ISO/IEC/IEEE 24765:2017 Systems and Software Engineering -- Vocabulary}~\cite{ISO24765} as our ground truth. This standard is a comprehensive reference containing 5,381 standardized definitions widely adopted in SE education and industry~\cite{SWEBOK4:2024}. We then construct falsified definitions from this ground truth using two complementary approaches. First, we use structural falsification by removing critical information from the definition of a term, while preserving the grammatical correctness of this. Such shortened versions are often casually used, though they lack the required precision to fully describe a term, e.g., because a qualifier or constraint is missing. Second, we use semantic falsifications where we systematically replace key terms within a definition with semantically similar but incorrect alternatives. This falsification approach reflects the kind of inadvertent word substitution that either results from misunderstanding or from overly casual use of terminology. We assess quantitatively by measuring the true positive rate for \textbf{H1} and the true negative rate for \textbf{H2}, respectively. Additionally, we qualitatively analyze reasoning traces to understand if these are coherent and consistent with the label assigned by the LLM to gain further insights into the its behavior and understand failure modes.

Surprisingly, \textbf{our results contradict both hypotheses}: all tested LLMs (Claude Opus 4.6, Claude Sonnet 4.6, GPT-5.2, GPT-5 Nano, and Gemini 2.5 Flash) are much better at identifying falsified definitions as problematic than identifying correct definitions as correct. This indicates a systematic bias within the LLMs to criticize terminology as incorrect, even when correctly used. Our analysis also reveal a certain rejection bias by the LLMs, i.e., most of the LLMs we study tend to look for, or even fabricate, reasons why definitions can be incorrect, rather than accepting them as correct. The outlier in our study is Opus 4.6, which is rather too lenient, i.e., it often recognizes that aspects are problematic, but accept definitions as correct anyway. Overall, the LLMs are unreliable judges of the correctness of definitions, which implies that they should be used with care whether consistent and correct usage of specific terms is important for a use case. 

The remainder of this paper is organized as follows. Section~\ref{sec:relatedwork} presents the background and related work on which we build. Afterwards, we define our research methodology in Section~\ref{sec:method}, including how we construct our definitions dataset, the prompt we use, how we measure the outcomes, and which LLMs we select and why. We proceed to present the results in Section~\ref{sec:results} and discuss them in depth in Section~\ref{sec:discussion}. From this, we derive some concrete and actionable implications for researchers and practitioners using LLMs in Section~\ref{sec:implications}. Finally, we discuss threats to the validity of our work in Section~\ref{sec:threats} and conclude in Section~\ref{sec:conclusion}.

\section{Background and related work}
\label{sec:relatedwork}

\subsection{Software Engineering Terminology}
\label{sec:relatedwork-terminology}

There are many standards related to SE. For example, the \textit{Software Engineering Body of Knowledge}~\cite{SWEBOK4:2024} provides a taxonomy of 36 standards grouping them into foundation, lifecycle processes \& concepts, assessment/governance, process elaborations, application guidelines, and artifact descriptions. Of these standards, the \textit{ISO/IEC/IEEE 24765:2017 Systems and Software Engineering -- Vocabulary}~\cite{ISO24765} (hereafter referred to as ISO 24765) serves a foundational role for the definition of harmonized SE terminology. The standard was jointly developed among standardization bodies and provides definitions of terminology based on related standards across the whole SE lifecycle. Indeed, the list of references in Annex A of ISO 24765 refers to 141 standards as source of the requirements. These standards come from different standardization bodies, i.e., the International Organization for Standardization (ISO), the International Electrotechnical Commission (IEC), the Institute of Electrical and Electronics Engineers (IEEE), and the Project Management Institute (PMI). As such, the ISO 24765 can be seen as the most authoritative source of the correct technical definition of SE terminology. 

Due to the central role and relationship to other standards, we use the ISO 24765 as the foundation for our work. However, there are still other definitions of terminology out there, e.g., the ISTQB Glossary~\cite{istqpglossary} or the CPRE Glossary~\cite{irebglossary}. The definitions provided by these standards are generally not in conflict with the ISO 24765 definitions, but may differ in concrete wording, specificity, or focus of the definitions. 

\subsection{Software Engineering Language Understanding Benchmarks}
\label{sec:relatedwork-selu}

Natural Language Processing (NLP) is often subdivided into Natural Language Understanding (NLU) and Natural Language Generation (NLG). Similarly, we can distinguish Software Engineering Language Understanding (SELU) from Software Engineering Language Generation (SELG). 

The most popular line of research for benchmarking is code-related SELG tasks, e.g., earlier benchmarks like HumanEval~\cite{humaneval}, MBPP~\cite{mbpp}, and EvalPlus~\cite{evalplus} that evaluate LLMs using synthetic, function-level Python tasks designed to assess basic coding capabilities. More recent benchmarks like SWE-bench~\cite{swebench}, BigCodeBench~\cite{bigcodebench}, and EvoEval~\cite{evoeval} source their data directly from realistic sources and aim to achieve a broader coverage of coding activities, with the recent benchmarks like DeepSWE~\cite{datacurveDeepSWE} and ProgramBench~\cite{yang2026programbench} focusing more and more on complex, long-context tasks. Moreover, there are also code-related SELU tasks like CodeXGlue~\cite{codexglue}. 

Since studies showed that software engineers spend a large amount of time on non-code activities~\cite{productivity1, productivity2}, there is also work that focuses specifically on non-code tasks. For example, the analysis of issue reports, e.g., to identify if the issue is a bug~\cite{bug_issue} or contains safety-related concerns~\cite{safety_issue}, analysis of requirements with respect to security aspects~\cite{security_requirement} or to understand if they are functional or non-functional~\cite{functional_requirement}, or to label SE terminology within texts through named entity recognition~\cite{se_entities}. The SELU benchmark~\cite{selu} integrates 22 such textual artifact-related tasks similar to NLU benchmarks such as GLUE~\cite{glue} and SuperGLUE~\cite{superglue}.

The tasks described above rely on a sound understanding of SE terminology, e.g., to differentiate security from safety, functional requirements from non-functional requirements, or have a clear notion of bugs. Nevertheless, we are not aware of any literature that directly studies this assumed underlying capability, i.e., the correct handling of SE terminology. 

\section{Methodology}
\label{sec:method}

\subsection{Dataset Construction}
\label{sec:dataset}

As discussed in Section~\ref{sec:relatedwork-terminology}, we use the ISO 24765 standard as our source for the definitions of SE terminology. Each entry for the 5,381 definitions that the standard defines consists of a term and one or more numbered definitions with references to the originating standard. We extract the textual definition and remove all metadata (e.g., references to other standards). This cleaning ensures that there are no unintentional cues for the LLM under evaluation. The extracted true definitions of terms give us the dataset $D_{true}$.

We then use $D_{true}$ as the foundation to develop a set with falsified definitions $D_{false}$, using two complementary falsification approaches to create plausible but incorrect definitions. Each approach targets a different dimension of definition integrity: \textit{structural falsification} removes critical semantic content, such that the resulting definitions remain syntactically coherent while containing meaningful inaccuracies that require domain understanding to identify; and \textit{semantic falsification} introduces subtle changes in meaning through lexical substitutions of words. A key design requirement across both methods is that the resulting falsified definitions must remain plausible. Falsifications that are too obvious, whether by removing too much content or by introducing clearly ungrammatical substitutions, would not reflect the real-world conditions the study aims to simulate and would trivialize the detection task. 

For each definition of a term $d_{true} \in D_{true}$, we apply the following steps for falsification:

\begin{enumerate}
    \item Conduct a structural falsification $d_{struct}$. If the falsification is valid, add $d_{struct}$ to $D_{struct}$ and terminate. Otherwise continue with step 2.
    \item Conduct a semantic falsification $d_{sem}$. If the falsification is valid, add $d_{sem}$ to $D_{sem}$. Otherwise stop without falsified version of $d_{true}.$
\end{enumerate}

Overall, this gives us $D_{false} = D_{struct} \cup D_{sem}$ as the set of falsified definitions such that every definition from the ISO 24765 standard appears only once in the falsified versions. The details for the falsification methods and their validity checks are described below. The choice to first apply the structural falsification is because this can be applied to a smaller subset of the overall set of definitions, since it requires a certain syntactic complexity of the definition, while the semantic falsification can be applied more broadly as it only requires the presence of a sufficient number of suitable individual words. Table~\ref{tbl:datasize} shows the overall size of the three datasets. When applied in this order, we get similar amounts of data for both falsification methods. For 763 definitions of terms, neither falsification method could be applied, e.g., because the definitions were too short. To have consistency between $D_{true}$ and $D_{false}$ we also drop these short definitions from $D_{true}$, such that we have a total of 4,618 definitions in both datasets.

\begin{table}[t]
\centering
\caption{Size of the data sets of true and falsified definitions}
\label{tbl:datasize}
\begin{tabular}{lcl}
\toprule
\textbf{Data set} & \textbf{Size} & \textbf{Description} \\
\midrule
$D_{true}$ & 5,381 & True definitions from ISO 24765 \\
$D_{struct}$ & 2,678 & Structurally falsified definitions \\
$D_{sem}$ & 1,940 & Semantically falsified definitions \\
$D_{false}$ & 4,618 & $D_{struct} \cup D_{sem}$ \\
\bottomrule
\end{tabular}
\end{table}

\subsubsection{Structural Falsification}\label{subsec:omission}

The structural falsification approach removes complete grammatical phrases that carry critical semantic information while preserving the overall sentence structure. The core idea is that technical definitions in the ISO 24765 standard rely heavily on qualifying phrases to distinguish a concept from related ones. Removing such a phrase produces a definition that remains grammatically complete and superficially plausible, but is now semantically underspecified in a way that requires domain knowledge to detect. To illustrate, consider the following definition of the term \textit{hyperparameter}:
\begin{quote}
\textit{Original:} variable used to define the structure of a neural network and how it is trained.
\end{quote}
\begin{quote}
\textit{Falsified:} variable used to define the structure and how it is trained.
\end{quote}
The structural falsification removes the prepositional phrase ``of a neural network''. The falsified definition remains grammatically correct and reads as a plausible general description of a variable, but has lost the critical qualifier that anchors it specifically to neural network configuration.

Using dependency parsing with spaCy~\cite{spacy}, we identify four types of deletable chunks: prepositional phrases, relative clauses, adjective modifiers, and appositives. These four types are specifically chosen because they function as modifiers or qualifiers that attach to the core of a sentence rather than forming the core itself. The subject, main verb, and primary object of a definition remain intact after deletion, ensuring the resulting falsified definition reads as a complete sentence. We prioritize chunk types in a fixed order heuristically by their expected semantic contribution to noun phrase interpretation~\cite{huddleston2002cambridge}:

\begin{enumerate}
    \item \textit{Prepositional phrases} are targeted first, as they frequently encode the scope or condition that makes a definition precise. For example, phrases such as ``based on specified requirements'' or ``in a primary element'' directly constrain the concept being defined.
    \item \textit{Relative clauses} are targeted second, as they similarly introduce constraints or qualifications on the concept being defined.
    \item \textit{Adjective modifiers} are targeted third, as they carry less critical information than \textit{prepositional phrases} or \textit{relative clauses} but can still alter the meaning of a definition when removed.
    \item \textit{Appositives} are used last, as they tend to carry the least essential information among the four chunk types.
\end{enumerate}

We only delete one phrase per definition. This is a deliberate design choice to keep the falsification subtle. If we would remove multiple phrases, this would increase the risk that definitions become obviously incomplete even to non-experts, which would undermine the goal of creating plausible falsifications that require genuine domain understanding to detect.

We also implement several safeguards to avoid that even deleting a single phrase leads to overly simple falsifications: 
\begin{itemize}
    \item We check if the deletion leaves the sentence ending in a grammatically incomplete state, e.g., due to a trailing bare preposition.
    \item We prohibit dropping critical relational phrases (e.g., ``in which'', ``of which'', ``by which'', and ``to which'') as this would break the syntactic structure of the sentence instead of simply omitting a qualifier.
    \item We do not apply this method at all to definitions with less than five words, since it is unlikely that the result (which must then be at most four words) is a sufficiently detailed definition for any term. 
\end{itemize}

\subsubsection{Semantic Falsification}
\label{subsec:embeddings}

The semantic falsification approach is based on the concept of replacing terms within a definition to change its semantic meaning while preserving the structure. The difficulty with this replacement is to select targets for replacement (i.e., which words within a definition should be replaced) and then identify the replacement candidates (i.e., words that would alter the meaning, but not in a drastic and trivial manner). The following example illustrates using the definition of the term \textit{ontology}:

\begin{quote}
\textit{Original:} logical structure of the \textbf{terms} used to describe a domain of \textbf{knowledge}, including both the definitions of the applicable terms and their relationships.
\end{quote}

\begin{quote}
\textit{Falsified:} logical structure of the \textbf{relation} used to describe a domain of \textbf{aspects}, including both the definitions of the applicable terms and their interaction.
\end{quote}
Each substitution is semantically plausible in isolation (relation instead of terms, aspects instead of knowledge), but within the context, this does not define an ontology anymore, since this is specifically about terms within a domain. Nevertheless, for non-experts this still seems plausible. 

Our targets for replacement are nouns longer than three characters, as these typically carry the most domain-specific meaning since they are usually content words and not function words~\cite{manning1999foundations}. Additionally, we predefined a set of technical adjectives commonly found in SE definitions as replacement targets, e.g., ``static'' or ``dynamic''. These adjectives carry precise technical meaning which makes them ideal candidates for replacements, since substitution breaks this meaning. 

We use GloVe word embeddings to identify replacement candidates~\cite{pennington2014glove}. In the same manner as other word embeddings, GloVe aligns dense numerical vectors in a continuous space such that semantically related words appear close to one another. We use terms that have a cosine similarity in the range of 0.55 to 0.70 to the target words as replacement candidates. Manual inspection found that candidates below 0.55 are too dissimilar and mostly introduced obviously wrong results, undermining plausibility. Candidates above 0.70 were often too similar, which leads to semantic-preserving replacements as they are effectively synonymous with the original. The 0.55 to 0.70 range therefore targets the region where words are semantically adjacent but not interchangeable. However, even within this range, we occasionally identified words that functioned as near-synonyms within the SE domain, or words that were semantically distant from the domain entirely. To address this, we first apply lemma filtering, i.e., we use spaCy's lemmatizer to exclude synonyms. Additionally, we iteratively build a blacklist of problematic replacements by manually inspecting the GloVe outputs and identifying recurring categories of inappropriate replacements, that the similarity range alone does not exclude.

In addition to this embedding-based replacement, we also identify a set of domain term pairs that we directly swap within the definition, e.g., substituting ``source'' with ``target'' or ``specification'' with ``implementation''. These pairs represent concepts that are closely related but technically distinct in SE, making a direct swap an effective falsification strategy without needing to consult the embedding space. Both the technical adjective set and domain term pairs were curated based on established SE domain knowledge by analyzing the definition of the terminology.

To further avoid problems caused by the replacement, we implement a set of safeguards that ensure that the replacements do not inadvertently break the definitions:

\begin{itemize}
    \item \textbf{Part-of-speech matching} ensures replacements preserve the grammatical structure of the original definition.
    \item \textbf{Context-aware checking} examines surrounding words to catch cases where a replacement is inappropriate given its immediate linguistic context, even if it passes the similarity and blacklist checks. Inappropriateness is determined to be the case if the similarity between the two words before and after the replacement target according to the GloVe embeddings is low.
    \item \textbf{Grammar checking} using the rules for inflection determines if the results introduce article-noun agreement errors or duplicate consecutive words, ensuring the falsified definition remains syntactically well-formed.
\end{itemize}

\subsection{Prompt}

All LLMs are evaluated in a zero-shot setting using a standardized simple chain-of-thought prompt, where the placeholder \texttt{[DEFINITION]} is replaced with the definition that should be assessed.

\begin{quote}
\texttt{Is this technical definition correct?}

\texttt{Definition: [DEFINITION]}

\texttt{Respond with True (correct) or False (incorrect) and explain your reasoning.}
\end{quote}

We collect the binary classification label ($true$ or $false$) and the accompanying explanation for each response. The binary response is extracted programmatically from the LLM output using pattern matching, while the explanation is preserved in full for subsequent reasoning quality analysis.

\subsection{Quantitative judgment analysis}\label{subsec:quantitative}

The main criterion for the evaluation of our hypotheses is whether the LLMs we test correctly determine the label. For $\textbf{H1}$ we measure the rate of correct predictions on $D_{true}$, which is effectively the true positive rate. Let $m(d)$ denote the label that the LLM assigns to a definition $d$. Then we have:

\begin{equation}
    TPR = \frac{|\{d \in D_{true}: m(d)=true\}|}{|D_{true}|}.
\end{equation}

Similarly, we assess how many falsified definitions are correctly identified as falsified for \textbf{H2}, which is effectively the true negative rate. We determine this both overall, as well as for the two falsification methods, i.e., we have:

\begin{align}
    TNR &= \frac{|\{d \in D_{false}: m(d)=false\}|}{|D_{false}|} \\
    TNR_{struct} &= \frac{|\{d \in D_{struct}: m(d)=false\}|}{|D_{struct}|} \\
    TNR_{sem} &= \frac{|\{d \in D_{sem}: m(d)=false\}|}{|D_{sem}|}.
\end{align}

\subsection{Qualitative reasoning analysis}
\label{sec:method-qualitative}

Beyond the quantitative metrics, we manually annotate LLM responses to assess the quality of reasoning behind each verdict. For this, we sample 100 instances from $D_{true}$ and 100 instances from $D_{false}$. For the results on correct definitions, we stratify this sampling by the correctness of the response, i.e., we sample 50 instances that are true positives, and 50 instances that are false negatives. The stratification for the falsified definitions is similar, but takes the falsification method into account. Hence, we sample 25 instances that are true negatives that were created with structural falsification, 25 instances that are true negatives that were created with semantic falsification, 25 instances that are false positives that were created with structural falsification, and 25 instances that are false positives that were created with semantic falsification. 

We use deductive coding with the following categories:

\begin{itemize}
    \item \textit{Correct answer, correct reasoning}: The LLM correctly accepts/rejects the definition and its justification is accurate and on-point.
    \item \textit{Correct answer, wrong reasoning:} The LLM correctly accepts/rejects the definition but its justification is inaccurate, misidentifies the concept, or points something out as incorrect, that is not the actual problem with a falsified definition or some fabricated reason.
    \item \textit{Wrong answer, correct reasoning:} For false negatives, this means that the LLM incorrectly rejects the definition, but its critique is coherent and accurate, meaning it simply applies a too strict standard to an authoritative definition. For false positives, this can be viewed as a lenient acceptance, i.e., the LLM engages with or notices the falsification but rationalizes it as acceptable and ultimately accepts the definition as correct. This indicates partial awareness of the error followed by self-persuasion into accepting it.
    \item \textit{Wrong answer, wrong reasoning:} For false negatives, this means that the LLM incorrectly rejects the definition based on a misread, an invented fault, or reasoning built on the wrong concept entirely. For false positives, this means the LLM completely misses the falsification with no indication of awareness and confidently accepts the definition as correct.
\end{itemize}

\subsection{LLM Selection}
\label{sec:models}

Our primary criteria for LLM selection are the capability tier and reasoning mode. The capability tier distinguishes between flagship LLMs and smaller, but cheaper alternatives. By selecting LLM with the same version from different capability tiers from the same vendor, we can judge if the performance differs between LLMs that were arguably trained using a similar approach (data, architecture). The reasoning mode is related to this, but in general a separate dimension. With this, we refer to LLMs where the generation of reasoning tokens is part of the post-training process~\cite{tie2025survey}. Since these reasoning tokens are typically hidden from or only summarized in the user-facing output, we still rely on an explicit final explanation of the reasoning even for such LLMs for our qualitative analysis described above. By pairing LLMs with and without reasoning from the vendor, and ideally from the same capability tier, we can assess whether reasoning affects the capability to understand terminology definitions. 

A secondary criterion is that we need to restrict the overall number of LLMs within our analysis, since the qualitative analysis requires linearly more human resources for judgment with each LLM. 

With these criteria in mind, we select the LLMs depicted in Table~\ref{tbl:models}. The three LLMs from OpenAI allow us to assess to which degree reasoning affects the results of a current flagship LLM, as well as how big the difference to a small LLM is. The two Anthropic LLMs give a similar perspective for a different LLM vendor, except that it is not possible to fully disable reasoning with the Opus LLM from Anthropic. Gemini-2.5-Flash-NR completes the selection with a third provider to increase the generalizability of our results. 

\begin{table}[t]
\caption{LLMs selected for the experimentation}
\label{tbl:models}
\centering
\begin{tabular}{lll}
\toprule
LLM & Description \\
\midrule
GPT-5.2-R & OpenAI's flagship LLM with reasoning \\
GPT-5.2-NR & OpenAI's flagship LLM without reasoning \\
GPT-5-Nano-NR & Smaller LLM from OpenAI without reasoning \\
Opus-4.6-R & Anthropic's flagship LLM with reasoning \\
Sonnet-4.6-NR & Smaller LLM from Anthropic without reasoning \\
Gemini-2.5-Flash-NR & Smaller LLM from Google without reasoning \\
\bottomrule
\end{tabular}
\end{table}

We acknowledge that an ideal experimental setup would include more LLMs, e.g., Haiku, also from Anthropic, more alternatives from Google or others open-weights LLMs like Qwen, GLM, Llama, or Kimi. However, given the six LLM we include, we already have to conduct $200 \cdot 6 = 1,200$ human judgments of the reasoning for our qualitative analysis (see Section~\ref{sec:method-qualitative}). This selection enable us to look at three vendors and LLMs of different types and to see both how capability tiers and reasoning affect results, while also being able to see patterns with different training regimes from vendors. Consequently, we believe that the selected LLMs give us a balanced trade-off between the costs associated with the high-effort qualitative analysis and the external validity of our results. 

\section{Results}
\label{sec:results}

\begin{figure}[h]
    \centering
    \includegraphics[width=\columnwidth]{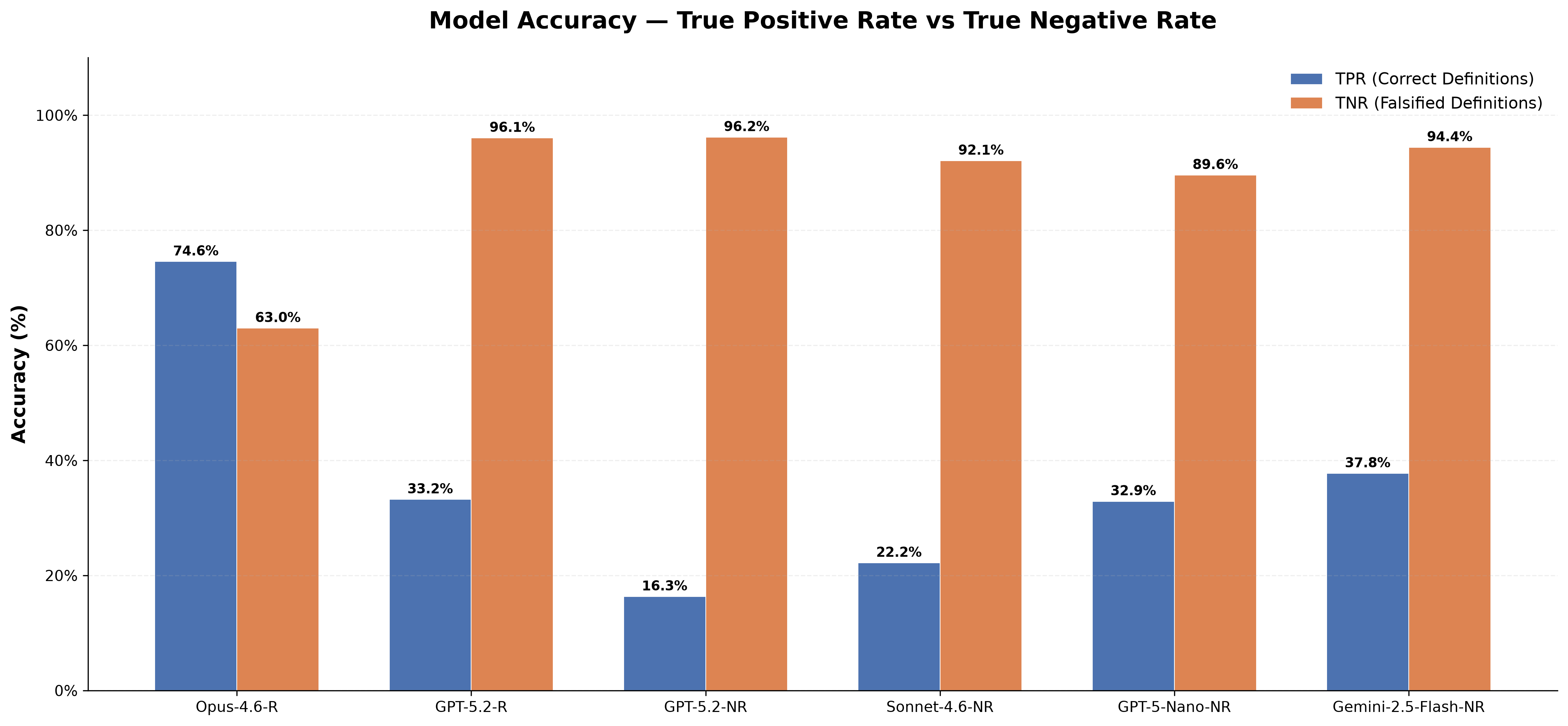}
    \caption{LLM accuracy -- True Positive Rate (TPR) on correct definitions versus True Negative Rate (TNR) on falsified definitions.}
    \label{fig:accuracy_overview}
\end{figure}

Figure~\ref{fig:accuracy_overview} summarizes our main results and shows that there is a mixed picture with respect to the capability of the LLMs to determine if definitions are correct. The TPR is generally a lot lower than the TNR, indicating that the LLMs are better at rejecting incorrect definitions than at accepting correct definitions. In the following, we dive into these results in greater detail and assess what this means for our hypotheses. 

\subsection{H1: Judgments on correct definitions}

Table~\ref{tbl:baseline} summarizes the results for the judgments on $D_{true}$. The LLMs are ranked by the best TPR. The data shows a surprisingly low quality in general as well as very large differences between the LLMs. Even the best LLM, Opus-4.6-R only achieves a TPR of 74.6\%, i.e., falsely labels about one quarter of correct definitions as wrong. The gap to the other LLM is huge: the second-best LLM Gemini-2.5-Flash only achieves a TPR of 37.8\%, and the worst LLM is GPT-5.2-NR with a TPR of only 16.3\%. In general, the reasoning LLMs from the same family outperform the non-reasoning LLMs. However, LLM size seems less relevant, as indicated by the performance of GPT-5-Nano-NR which is very close to the much larger GPT-5.2-R. 

Table~\ref{tbl:baseline} also gives us further insights into the reasoning quality based on a sample of 100 responses per LLM as described in Section~\ref{sec:method-qualitative}. As can be expected, the reasoning quality on the correct predictions is generally relatively high (columns CC/CW), with GPT-5.2-R even perfectly reasoning for all correct predictions and Gemini-2.5-Flash-NR with mistakes for only 2\%.\footnote{Since we have 50 true positives in the samples, this means we have a single mislabel.} The other four LLMs make mistakes for 16\%-20\% of the correct predictions. The reasoning quality on false negatives is generally poor: the reasoning contained fabricated rather than actual faults in the definitions, which shows that the bad labeling is indeed due to a lack of SE terminology understanding. The outlier is GPT-5.2-R, which actually reasons coherently and accurately, but then applies a too strict standard for terminology that requires more than is actually required for a correct definition.

\begin{figure}[h]
    \centering
    \includegraphics[width=\columnwidth]{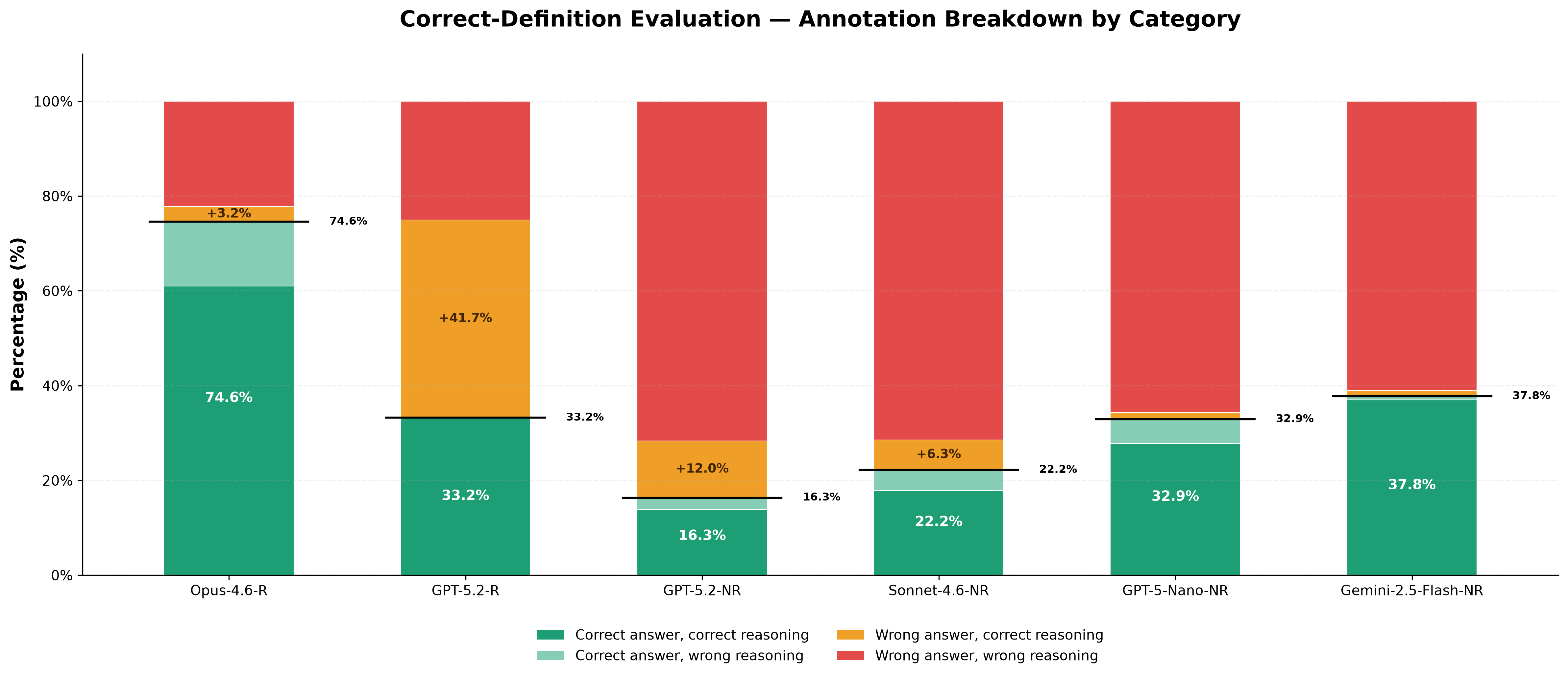}
    \caption{Reasoning quality breakdown for $D_{true}$. Values are estimated based by extrapolating from the TPR and the CC/CW ratios, and the FNR and the WC/WW ratios reported in Table~\ref{tbl:baseline}.}
    \label{fig:baseline_reasoning}
\end{figure}

When we extrapolate the reasoning correctness from the sample to the full data, we get the breakdown of results shown in Figure~\ref{fig:baseline_reasoning}. The stacked bars highlight the joint perspective between the TPR, FNR, and the qualitative labels: most results are clear mistakes. Opus-4.6-R has a decent, though far from perfect TPR, but this is partially because of wrong reasoning. If all cases where GPT-5.2-R reasoned correctly had gotten the correct label as well, the performance would be comparable to Opus-4.6-R. 

\begin{mdframed}
\textbf{Our results consistently and clearly contradict H1, i.e., all LLMs have problems identifying correct definitions as correct. The flagship LLMs Opus-4.6-R and GPT-5.2-R show the best performance, but still fail clearly on at least one quarter of the SE terms.} 
\end{mdframed}

\begin{table*}[t]
      \caption{LLM accuracy on correct definitions $D_{true}$. TP are the true positives, TPR the true
    positive rate, FN the false negatives, and FNR the false negative rate. The columns CC, CW, WC, WW
    summarize the qualitative insights into the reasoning correctness: the first letter means label is
    (C)orrect/(W)rong, second letter means reasoning is (C)orrect/(W)rong.}
      \label{tbl:baseline}
      \centering
      \begin{tabular}{lrrrr|rrrr}
      \toprule
      LLM & \#TP & TPR & CC & CW & \#FN & FNR & WC & WW \\
      \midrule
      Opus-4.6-R          & 3,444 & 74.6\% & 80\% &  20\% & 1,174 & 25.4\% &  12\% & 88\% \\
      Gemini-2.5-Flash-NR & 1,743 & 37.8\% & 98\% &  2\% & 2,874 & 62.2\% &  2\% & 98\% \\
      GPT-5.2-R           & 1,534 & 33.2\% & 100\% &  0\% & 3,084 & 66.8\% & 62\% & 38\% \\
      GPT-5-Nano-NR       & 1,519 & 32.9\% & 84\% &  16\% & 3,099 & 67.1\% &  2\% & 98\% \\
      Sonnet-4.6-NR       & 1,025 & 22.2\% & 80\% & 20\% & 3,593 & 77.8\% &  8\% & 92\% \\
      GPT-5.2-NR          &   755 & 16.3\% & 84\% &  16\% & 3,863 & 83.7\% &  14\% & 86\% \\
      \bottomrule
      \end{tabular}
    \end{table*}

\subsection{H2: Judgments on falsified definitions}

Table~\ref{tbl:falsified} shows the results on the falsified definitions $D_{false}$. Here, the LLMs generally have a very strong performance with a TNR between 89.6\% and 96.2\%, with the exception of Opus-4.6-R that only achieves a TNR of 63.0\%. Whether LLMs have reasoning enabled or not does not seem to be a driving factor here, highlighted by the almost equally strong performance of GPT-5.2-R and GPT-5.2-NR. LLM size yields no reliable tendencies: GPT-5-Nano-NR is weaker than the GPT-5.2 LLMs, but Sonnet-4.6-NR clearly outperforms Opus-4.6-R. 

In line with the generally high performance, the reasoning is also typically correct, highlighted by the high values for CC reported in Table~\ref{tbl:falsified}. The mistakes are also mostly based on faulty reasoning. The clear exception here is again Opus-4.6-R, which often reasons correctly regarding mistakes, but still accepts them as correct. A detailed analysis regarding the reasons for the deviations of Opus-4.6-R from the other LLMs can be found in the Discussion in Section~\ref{sec:discussion}.

\begin{figure}[h]
    \centering
    \includegraphics[width=\columnwidth]{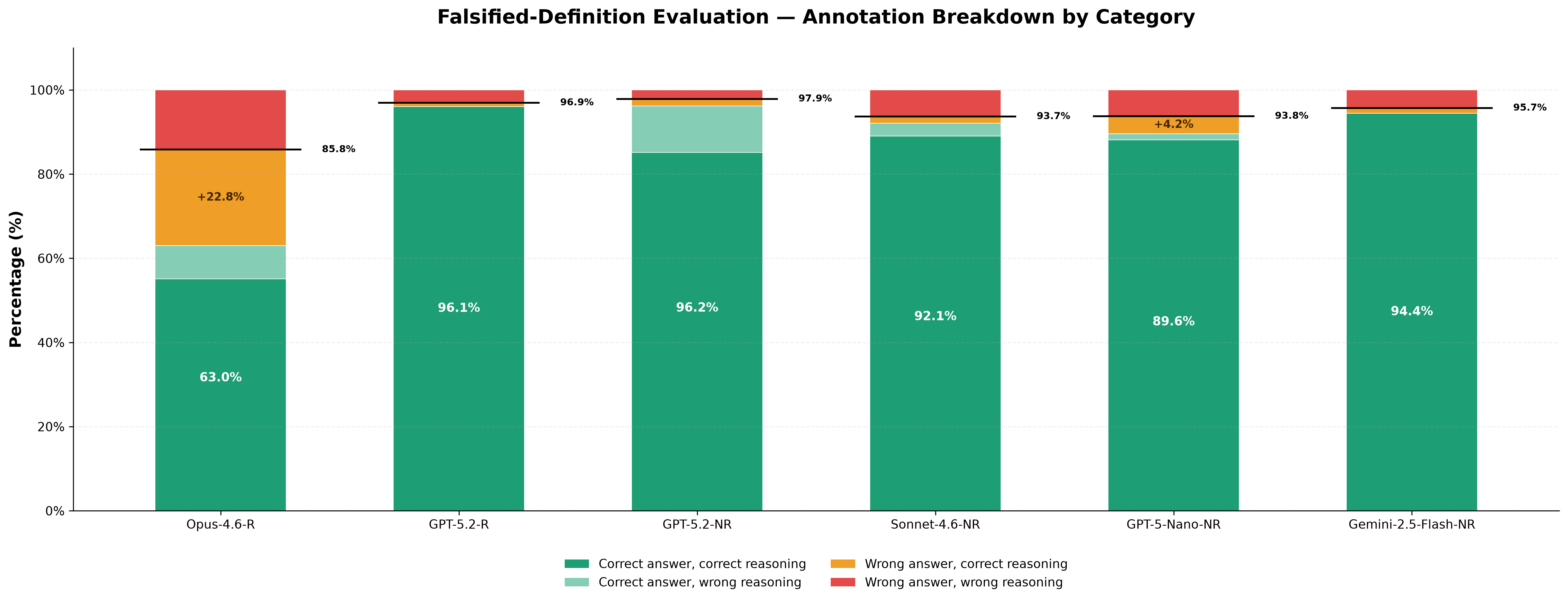}
    \caption{Reasoning quality breakdown for $D_{false}$. Values are estimated based by extrapolating from the TNR and the CC/CW ratios, and the FPR and the WC/WW ratios reported in Table~\ref{tbl:falsified}.}
    \label{fig:falsified_reasoning}
\end{figure}

Same as before, we extrapolate the reasoning correctness from the sample to the full data and get the breakdown of results as shown in Figure~\ref{fig:falsified_reasoning}. This highlights what we observed before: all LLMs are very strong and reasoning is mostly a neutral factor, except for Opus-4.6-R. If the LLM went through with the reasoning and not leniently accept definitions with recognized flaws as nevertheless correct, the performance would be closer to the other LLMs, but still not at the same level. 

\begin{table*}[t]
    \caption{LLM accuracy on falsified definitions $D_{false}$. TN are the true negatives, TNR the true
  negative
    rate, FP the false positives, and FPR the false positive rate. The columns CC, CW, WC, WW
    summarize the qualitative insights into the reasoning correctness: the first letter means label is
    (C)orrect/(W)rong, second letter means reasoning is (C)orrect/(W)rong.}
    \label{tbl:falsified}
    \centering
    \begin{tabular}{l rrrr|rrrr}
    \toprule
    LLM & \#TN & TNR & CC & CW & \#FP & FPR & WC & WW \\
    \midrule
    GPT-5.2-NR          & 4,442 & 96.2\% & 88\% &  12\% &   176 &  3.8\% & 44\% & 56\% \\
    GPT-5.2-R           & 4,436 & 96.1\% & 100\% &  0\% &   182 &  3.9\% &  22\% & 78\% \\
    Gemini-2.5-Flash-NR & 4,361 & 94.4\% & 100\% &  0\% &   257 &  5.6\% &  22\% & 78\% \\
    Sonnet-4.6-NR       & 4,252 & 92.1\% & 96\% &  4\% &   365 &  7.9\% &  20\% & 80\% \\
    GPT-5-Nano-NR       & 3,902 & 89.6\% & 98\% &  2\% &   453 & 10.4\% & 40\% & 60\% \\
    Opus-4.6-R          & 2,908 & 63.0\% & 88\% &  12\% & 1,708 & 37.0\% & 62\% & 38\% \\
    \bottomrule
    \end{tabular}
    \end{table*}
    
All LLMs achieved higher detection rates on semantic falsifications than on structural ones (see Table~\ref{tbl:method}). Indeed, the semantic falsifications are detected with a TNR of 95.5\% or higher, with GPT-5.2-R even reaching 99\%. The exception is Opus-4.6-R, which only reaches a TNR of 83.2\%. This means that LLMs are very good at spotting terms that are similar to the required term and identifying wrongly used terminology. While the rate for the detection of structural falsifications is still 87\% or higher, except for Opus-4.6-R which only achieves a TNR of 48.4\%, it seems like detecting that something crucial is missing from the context is more difficult for the LLMs. The possible reason for this can be seen in the correct reasoning with wrong results by Opus-4.6-R: the LLM frequently detected that something was missing, including what was missing, but judged that the definition of the term was acceptable nevertheless. Thus, while using inappropriate terms is a clear case for separation between correct and false definitions, missing details seem to be a fuzzier criterion. 

\begin{table}[t]
\caption{TNR by falsification method}
\label{tbl:method}
\centering
\begin{tabular}{lrr}
\toprule
LLM & $TNR_{Struct}$ & $TNR_{Sem}$ \\
\midrule
Opus-4.6-R          & 48.4\% & 83.2\% \\
GPT-5.2-R           & 93.9\% & 99.0\% \\
GPT-5.2-NR          & 94.4\% & 98.7\% \\
Gemini-2.5-Flash-NR & 91.6\% & 98.3\% \\
Sonnet-4.6-NR       & 89.6\% & 95.5\% \\
GPT-5-Nano-NR       & 87.2\% & 96.0\% \\

\bottomrule
\end{tabular}
\end{table}

\begin{mdframed}
\textbf{Our results consistently and clearly contradict H2, i.e., all LLMs are highly capable of recognizing the mistakes we introduced into definitions, though -- interestingly -- the arguably strongest LLM in our benchmark Opus-4.6-R makes most mistakes since it often recognizes mistakes, but accepts definitions nevertheless.} 
\end{mdframed}

\section{Discussion}
\label{sec:discussion}

For both hypotheses, our results contradict our expectations meaning that LLMs clearly do not understand SE terminology at an expert level, at least not natively through their weights without providing more information through context files or Web search. Within this section, we further explore possible reasons for this and then update the theory from which we started, to account for these new results. 

\subsection{Rejection Bias and the Opus Exception}\label{subsec:rejection-bias}

A key aspect of our evaluation design is that we test on both correct definitions $D_{true}$ and falsified definitions $D_{false}$. While we discuss the results separately above, in line with our hypotheses, a trend towards rejection emerges when we consider both together. This means that there seems to be a general tendency by the LLMs to judge definitions as incorrect. While our study design is not suitable to reliably identify the underlying mechanism for this, the qualitative results hint at the likely reasons: the LLMs tend to fabricate/hallucinate information in the sense that they either misinterpret terms or make up additional qualifiers that a definition requires. This interpretation is supported by the frequently occurring wrong reasoning on the false negatives. Through this tendency, the LLMs default to a rejection behavior, which automatically also explains the strong performance in detecting falsified definitions. GPT-5.2-NR illustrates this pattern most clearly. It achieved the highest detection rate of 96.2\% on falsified definitions, yet simultaneously rejected 83.7\% of correct definitions, the highest rejection rate of all LLMs.

Opus-4.6-R stands apart from this pattern. With a TPR of 74.6\%, it is the only LLM that correctly accepted the majority of correct definitions, suggesting that it does not exhibit the same systematic rejection bias by fabricating reasons for rejection. Viewed through this lens, its TNR of 63\% on falsified definitions is not straightforwardly a sign of poor detection capability. Rather, it may reflect a LLM that is less prone to systematic rejection, accepting definitions that appear plausible regardless of whether they are correct or not. Indeed, Opus-4.6-R rather seems to have an acceptance bias, as can be seen on the falsified definitions $D_{false}$ that Opus-4.6-R did often correctly engage with the definitions, but still opted to accept as correct.

\subsection{Impact of reasoning}\label{subsec:rq2}

The reasoning quality annotation reveals that correct detection does not always imply correct reasoning, i.e., the LLMs sometimes come to the correct results based on faulty assumptions. This disconnect between verdict accuracy and reasoning quality holds for both the baseline and falsified definition evaluations. If we were to use stricter criteria for correctness, i.e., require both a correct answer and a correct explanation of the answer, the performance of all LLMs would be weaker, though most LLMs would still perform well at identifying falsified definitions. However, as can possibly be expected, it seems like using LLMs where reasoning is part of the post-training is beneficial for the reasoning correctness over a normal chain-of-thought setting, as is demonstrated using the pairing of GPT-5.2-R and GPT-5.2-NR. Enabling reasoning in GPT-5.2-R did not meaningfully improve detection of falsified definitions (96.1\% versus 96.2\% for GPT-5.2-NR). However, GPT-5.2-R reasoned almost always correctly, while GPT-5.2-NR made some mistakes here. A similar difference can also be observed on $D_{true}$. 

\subsection{Detection Performance by Falsification Method}\label{subsec:performance-falsification-method}

All LLMs detected semantic falsifications more reliably than structural ones. This asymmetry is intuitive, considering how the LLMs are trained. Semantic falsifications introduce a term that is plausible without knowing the context based on the GloVe word embeddings, whereas the modern LMs are directly trained to identify how good terms fit within a context through the Causal Language Modeling (CLM) objective. Thus, one may argue that semantic falsification plays to the strengths of the LLMs, which explains that such falsifications are detected so reliably. 

Structural falsifications, by contrast, produce a definition that is incomplete but not internally contradictory. Nothing in the text is technically wrong. It is simply missing something. Detecting this requires the LLM to compare what is presented against what a complete definition should contain. Indeed, this is never part of training which just focuses on what comes next. To confirm that this is the case, future experiments could explore a falsification method that adds plausible content to falsify the definition. 

\subsection{Failure modes}

To better understand consistent failure modes, we looked for definitions that were problematic for all LLMs. Overall, we found 69 falsified definitions where all LLMs missed the problem. Of these definitions, 62 (89.9\%) were structural. This suggests that certain structural falsifications are fundamentally difficult to detect regardless of LLM scale or capability, e.g., cases where the removed information is sufficiently peripheral, or the remaining text sufficiently plausible, that no LLM flags it as incomplete. As an example, consider the structural falsification of the definition of \textit{conversational}:

\begin{quote}
\textbf{Original:} pertaining to an interactive system or mode of operation in which the interaction between the user and the system resembles a human dialog.
\end{quote}

\begin{quote}
\textbf{Falsified:} pertaining to an interactive system or mode of operation in which the interaction resembles a human dialog.
\end{quote}

The falsification omits the qualifier ``between the user and the system''. This part is certainly crucial for the definition, since this could otherwise also be the interaction between two systems. Nevertheless, what remains is plausible, especially considering the general meaning of the term ``conversation'' as a form of communication between human actors. Picking up such subtleties would certainly also be difficult for humans who are asked to judge the correctness of definitions. 

We further identified 78 cases where both reasoning LLMs accepted a falsified definition while at least one non-reasoning LLM correctly rejected it. These cases help us understand why reasoning may not only be neutral, but actually problematic for this task. To understand what happened, consider the definition of \textit{standby redundancy} and its structural falsification:

\begin{quote}
\textbf{Original:} in fault tolerance, the use of redundant elements that are left inoperative until a failure occurs in a primary element.
\end{quote}

\begin{quote}
\textbf{Falsified:} in fault tolerance, the use of redundant elements that are left inoperative until a failure occurs.
\end{quote}

The falsified version omits the phrase ``in a primary element'', leaving the definition to state only that the standby element is activated when a failure occurs, i.e., we have something that is grammatically complete but semantically underspecified.

Both Opus-4.6-R and GPT-5.2-R accepted the falsified definition. GPT-5.2-R's response is particularly revealing: in its own reasoning, it paraphrased the definition as the standby element being activated upon a failure of the primary element, effectively reinstating the omitted phrase from its own background knowledge, while simultaneously classifying the incomplete definition as correct. Rather than detecting the gap, the LLM filled it. This points to a specific failure mode in reasoning-capable LLMs -- and possibly very large LLMs -- in general: their breadth of background knowledge allows them to complete an underspecified definition, masking the very omission that makes it incorrect. Non-reasoning LLMs, which evaluate the definition more literally, instead flag the incompleteness. In this context, richer background knowledge worked against detection rather than in favor of it.

A related but distinct pattern appears in cases where the LLM explicitly notices a potential issue in its reasoning but still accepts the definition. The cases here involve the LLM surfacing relevant information in its reasoning trace and then dismissing it. The definition of \textit{demand paging} provides a clear instance of this pattern:

\begin{quote}
\textbf{Original:} storage allocation technique in which pages are transferred from auxiliary storage to main storage only when those pages are needed.
\end{quote}

\begin{quote}
\textbf{Falsified:} storage allocation technique in which pages are transferred from auxiliary storage only when those pages are needed.
\end{quote}

The falsified version omits ``to main storage'', leaving out the destination of the page transfer. Opus-4.6-R accepted the definition while its reasoning described pages being loaded ``into main memory only when they are referenced'', reinstating the missing element in its own words while rating the incomplete definition as correct.

In this case, the LLM's reasoning contained the information needed to identify the falsification, but that information did not influence the final classification. This suggests that in certain cases the reasoning trace and the classification decision operate somewhat independently, with the LLM's prior confidence in a well-known definition overriding what its own analysis surfaces.

\subsection{An updated theory for the definition understanding of LLMs}

Recall that we initially had the theory that LLMs are very good at identifying correct definitions, which led to the prediction of \textbf{H1}, but also that they might have problems with identifying falsification due to data diversity, which led to the prediction of \textbf{H2}. As is common with the scientific method, whenever observations disagree with the theory, the theory needs to be adjusted. Based on our results and the insights we discuss above, we instead adjust our theory as follows:
\begin{mdframed}
\textbf{Theory}: \textit{While LLMs have strong knowledge about SE terminology, they only have a superficial understanding of how different terms and concepts relate to each other. While this allows LLMs to identify if terms are used within the wrong context, they fail to reliably judge definitions of terms, where the contexts are correct. When the contexts are incomplete, they have problems recognizing this, as demonstrated by the problems with structural falsifications. When the whole context is correct, they make wrong connections between contexts, leading to the rejection of correct definitions due to hallucinations. These properties are a direct result of the pre-training objective (i.e., CLM), which reinforces syntactic co-occurrence relationships instead of directly training semantic relationships.}
\end{mdframed}

This theory is in line with our observations, including all failure modes we observe. It also predicts what should happen with other types of falsification, e.g., that adding irrelevant, but plausible content should be hard for the LLM to detect, if it matches the general semantics of the context. However, if irrelevant out-of-context information is added, LLMs should detect this. Future work should design suitable experiments to further test, and possibly refine, this theory. 

\section{Implications}
\label{sec:implications}

Our study has further implications for anybody using LLMs for SE tasks, both researchers and practitioners, beyond the immediate, possibly general problem with LLMs that our theory suggests. Importantly, whenever LLMs are used for tasks that do require an in-depth understanding of specific terminology, one cannot rely on the LLM to provide this knowledge. This means that instead of using an implicit prompt that does not define the clear semantics of a term, \textit{important terminology should be explicitly defined to avoid subtle misunderstandings} or possible mistakes. Of course, this presumes that if the correct definitions are supplied, the background knowledge of the LLMs will be sufficient to apply them within a context. Notably, this is nothing new within SE: requirements specifications contain glossaries for exactly this purpose and the very ISO 24765 standard was developed to facilitate this common terminology. The takeaway should thus be that while LLMs are often powerful, they still have weaknesses, some of them similar to human weaknesses. Thus, when defining an agent for a task or when using an LLM to judge SE data, \textit{we need to make sure to make our language explicit and specific}, to avoid misinterpretations and, thereby, downstream mistakes. Our results show that \textit{this does not only mean project-specific terminology, but also the SE terminology itself}. 

Further, specifically for researchers, \textit{we caution against the use of LLMs for data annotation in contexts where information is provided implicitly}. Our results show that while the LLMs can reason about underspecified information (in our case, the structural falsification), they are not reliable in this aspect: they are prone to misjudge the extension of the underspecified terms, and even when they get this right, they can fail to take this correctly into account. 

\section{Threats to validity}
\label{sec:threats}

There are several threats to the validity of our study. Importantly, we only studied definitions from the ISO 24765 standard. While the standard is comprehensive and unifies definitions from different sources, there are alternative definitions of terms available, including other standards as discussed above. While we see no hint towards this within our data, we cannot rule out that our results would look different, if other standards were considered. Importantly, we found no hint that the LLMs were preferring definitions from other standards, since most mistakes were genuine reasoning issues that cannot be explained by alternative definitions. Similarly, while we found consistent results across LLMs in general, the somewhat different behavior of Opus-4.6-R could indicate that our results might not generalize to other LLMs. Our qualitative analysis was based on 200 manually annotated examples per LLM. We cannot rule out human biases during this annotation, though we provided a clear specification for the deductive labels we used. We also only evaluated all LLMs using a single prompt template with no variation in phrasing. While recent LLMs have become more robust towards specific wording issues, we cannot rule out that this hindered the performance of some LLMs. However, experiments with more variants would have linearly increased the effort for the manual annotation, which is why we opted against this. 

Furthermore, our initial theory was derived from two assumptions about LLM training and not from existing empirical evidence. While we believe that we had a reasonable foundation for our hypotheses, one might challenge that the results should actually be expected as they were. However, even if this were the case, this does not alter the results at which we arrived, i.e., the contradictions of our hypotheses. Finally, while we derive a new theory based on the properties we observe in our results, we have not yet tested this theory, i.e., this theory should be seen as exploratory and not yet confirmed. An alternative theory could also be that the LLMs just generally have a rejection bias and all other findings are constructed by our active search for failure modes, instead of genuine properties of the data. Still, this is how theory is developed and future experiments need to provide evidence agreeing or contradicting our theory, to further advance our knowledge in this regard.

\section{Conclusion}
\label{sec:conclusion}

We evaluated how good six LLMs are at the task of understanding SE definitions. Contrary to our belief, we found that they struggle to identify definitions as correct and are much better at detecting mistakes within definitions. Our findings suggest a rejection bias within LLMs, which we attribute to a lack of in-depth understanding of connections between terms. Instead, we believe that the LLMs rather learn superficial relationships between terms, that allow them to spot certain mistakes (e.g., out-of-context usage of terms), but make them weak at judging if all required information is present. This leads to mistakes either through fabricating aspects that are supposedly required or, vice versa, failing to detect that crucial aspects are missing. Future work should further investigate this phenomenon, because the lack of principled understanding of SE terminology can have negative effects on any SE task that requires clear adherence to the specific meaning of well-defined terms. 

The data, prompts, scripts, and detailed results can be found in our replication kit~\cite{replication-kit}.

\bibliographystyle{IEEEtran}
\bibliography{refs}

\end{document}